\documentclass[%
reprint,
superscriptaddress,
showpacs,preprintnumbers,
amsmath,amssymb,
aps,
pra,
]{revtex4-1}

\pdfoutput=1

\usepackage{graphicx}
\usepackage{dcolumn}
\usepackage{bm}

\begin{document}

\title{Thermalized polarization dynamics of a discrete optical waveguide system \\ with four-wave mixing}

\author{S.A. Derevyanko}
 \email{stanislav.derevyanko@weizmann.ac.il}
 \altaffiliation[Also at ]{Aston Institute of Photonic Technologies, Aston University, UK}
\affiliation{%
 Department of Physics of Complex Systems, Weizmann Institute of Science, Rehovot 76100, Israel
}%


\date{\today}

\begin{abstract}
Statistical mechanics of two coupled vector fields is studied in the tight-binding model that describes propagation of
polarized light in discrete waveguides in the presence of the four-wave mixing. The energy and power conservation laws enable the formulation
of the equilibrium properties of the polarization state in terms of the Gibbs measure with positive temperature.
The transition line $T=\infty$ is established beyond which the discrete vector solitons are created. Also in the limit of the large
nonlinearity an analytical expression for the distribution of Stokes parameters is obtained which is found to be dependent only on the statistical properties of the initial polarization state and not on the strength of nonlinearity. The evolution of the system to the final equilibrium state is shown to pass through the intermediate stage when the energy exchange between the waveveguides is still negligible. The distribution of the Stokes parameters in this regime has a complex multimodal structure strongly dependent on the nonlinear coupling coefficients and the initial conditions.

\end{abstract}

\pacs{05.45.-a, 42.65.Wi, 42.25.Lc}
\maketitle


\section{\label{sec:intro} Introduction}
The equilibrium behavior and the equipartition of energy between various degrees of freedom in nonlinear, nonintegrable discrete systems has attracted considerable interest since the seminal study of Fermi, Pasta and Ulam \cite{fpu74}. In Hamiltonain systems with conserved number of excitations (waves) the maximum entropy principle suggests that in the final state of thermal equilibrium the statistics of the system is given by grand canonical Gibbs distribution with the effective ``temperature'' and ``chemical potential'' \cite{lrs88,jtz00,rckg00,EarlyRumpf,jr04}. However unlike the conventional statistical mechanics the effective temperature of this grand-canonical distribution depends on the initial position in the phase space and for certain regions can become negative making the distribution non-normalizable. Such regime is commonly attributed to the emergence of stable, localized, nonlinear structures corresponding to solitons in continuous systems \cite{jtz00} and discrete breathers \cite{rckg00,r08,r09} in discrete systems. From the point of view of the wave turbulence theory \cite{zlf,nnb01} the resulting equilibrium distribution provides stationary Rayleigh-Jeans spectra \cite{zlf}.
Also thermalization of light in nonlinear multimode waveguides and cavities has recently attracted attention in the context of classical optical wave condensation \cite{Picozzi}.

Here we will study the phenomenon of thermalization in the context of light propagation in a system of coupled nonlinear optical waveguides but the results can have wider applicability beyond the scope of the nonlinear optics. When the individual waveguide modes are strongly localized the nonlinear propagation of light is most commonly modelled by the discrete nonlinear Schrodinger equation (DNLSE) \cite{lscasy08}. In fact most studies of thermalization in nonlinear \textit{discrete} systems have concentrated on DNLSE in one  \cite{rckg00,EarlyRumpf,jr04,r08,r09,slbsm09} or two \cite{sps11} dimensions. Thanks the plethora of the results in the field of ``DNLSE thermalization'' the structure of the final equilibrium state and the thermodynamical conditions for the occurrence of discrete breathers are now well understood. Among numerous discoveries in this area we would like to point reader's attention to the universal correlations in 1D systems of optical waveguides predicted in \cite{slbsm09} in the limit when the nonlinearity dominates over the linear coupling. In this limit the effective dimensionless temperature turns out to be a universal constant independent on system parameters (provided that the initial state is characterised by uniform intensities) and the same universality is also manifested in the shape of the field correlation function.

In this paper we would like to focus on a much less studied model, namely, the thermalization of two coupled fields in the presence of the four-wave mixing (FWM) \cite{lscasy08}. In the context of nonlinear optics the situation corresponds to the propagation of polarized light in the birefringent material \cite{mhcssma03,hkw06} or mode interaction from different Floquet-Bloch bands \cite{mesma03}. Here we will concentrate on the first case, however the results presented here are quite general and can be applied to other nolinearly coupled systems. In order to give reference to the real-world units we use AlGaAs as an common example of a material with cubic symmetry and fused silica as the corresponding example of isotropic crystal.

\section{The model}
The wave dynamics of the two orthogonally polarized fields is given by the following pair of coupled equations \cite{mhcssma03,hkw06}:
\begin{subequations}
\label{vector-dnlse}
\begin{eqnarray}
i\,\frac{d \,a_n}{dz} &+& k \,a_n + C(a_{n-1}+a_{n+1}) +\gamma(|a_n|^2+\lambda_1\,|b_n|^2)a_n \nonumber \\
& + & \gamma \,\lambda_2\,b_n^2 \,a_n^* =0 \\
i\,\frac{d \,b_n}{dz} &-& k \,b_n + C(b_{n-1}+b_{n+1}) +\gamma(|b_n|^2+\lambda_1\,|a_n|^2)b_n \nonumber \\
& + & \gamma \,\lambda_2\,a_n^2 \,b_n^*=0, \quad n=1,\ldots, N
\end{eqnarray}
\end{subequations}
In the above equations, $a_n$ and $b_n$ are slowly varying filed envelopes of the TE and TM polarized waves, $k=k_0(n_x-n_y)/2$ is the polarization mode dispersion constant, $k_0$ is the vacuum wave vector, $n_x-n_y$ is the linear birefringence ($n_x-n_y=1.8\times 10^{-4}$ for AlGaAs), $C$ is the coupling constant ($C \sim 0.1$ mm$^{-1}$), $\gamma=(k_0/2) n_2 n (\varepsilon_0/\mu_0)^{1/2}$ is the nonlinear coefficient, $n_2$ is the Kerr coefficient ($n_2=1.5 \times 10^{-13}$ cm$^2$/W for AlGaAs), $n=n_x$ is the linear refractive index ($n=3.3$ for AlGaAs). The dimensionless constants $\lambda_1$ and $\lambda_2$ represent the relative strength of self- and cross-phase modulation (SPM and XPM). If one puts $\lambda_1=\lambda_2=0$ the system (\ref{vector-dnlse}) breaks into two independent scalar DNLSE equations. We can restrict ourselves to the case of positive coupling $C>0$ since the case of negative coupling can be recovered via a standard staggered transformation $a_n \to (-1)^n a_n$, $b_n \to (-1)^n b_n$. The change of sign in the nonlinearity can be also be compensated via more complicated transformation which involves staggering, complex conjugation and swapping: $a_n \to (-1)^n b_n^*$, $b_n \to (-1)^n a_n^*$. Both transformations only affect the field correlation functions and phase distributions (and not e.g. the intensity distributions) in a controlled way and here without loss of generality we will also restrict ourselves to the case of positive nonlinearity. In this paper we assume periodic boundary conditions although in the thermodynamic limit $N \to \infty$ this choice is not essential. Note in passing that continuous analogues of system (\ref{vector-dnlse}) were studied in \cite{contvectdnls} with regards to pulse propagation in optical fibers.

In any chosen nonlinear medium the dimensionless XPM and FWM constants, $\lambda_1$ and $\lambda_2$, are not independent and 3 possible cases of interest can be envisaged \cite{aa}:
\begin{enumerate}
\item[(a)] Anisotropic cubic medium (e.g. AlGaAs): $\lambda_1=2\lambda_2$.
\item[(b)] Generic isotropic medium: $\lambda_1 = 1-\lambda_2$.
\item[(c)] Isotropic cubic medium (e.g. fused silica): $\lambda_1=2/3$, $\lambda_2=1/3$.
\end{enumerate}
We will refer to cases (a) and (b),(c) as isotropic and anisotropic correspondingly.

The system is Hamiltonian with the Hamiltonian:
\begin{widetext}
\begin{equation}
\label{Hamiltonian-vect}
\mathcal{H} =k\sum_n (|a_n|^2-|b_n|^2) + C \sum_n (a_na_{n+1}^* +b_nb_{n+1}^* +c.c.)
 + \frac{\gamma}{2}\, \sum_n\left(|a_n|^4+|b_n|^4+2\lambda_1|a_n|^2|b_n|^2+2 \lambda_2 \mathrm{Re}(a_n^2 {b_n^*}^2)\right),
\end{equation}
\end{widetext}
which is a natural conserved quantity in the system while the additional integral of motion is provided by the total pulse power (proportional to the sum of local intensities)
\[
\mathcal{P}=\sum_n (|a_n|^2+|b_n|^2).
\]
Additionally in the absence of the four-wave mixing ($\lambda_2=0$) the individual powers in each polarization $P_a=\sum_n |a_n|^2$ and $P_b=\sum_n |b_n|^2$ are conserved.

In the state of thermodynamic equilibrium the stationary field distribution $P(\{a_n,a_n^*,b_n,b_n^*\})$ maximizes the entropy
$S=-\int P \,\ln P \,\prod_n da_n \, da^*_n \, db_n \, db^*_n$. However the nonlinear evolution always takes place on the shell $\mathcal{H}=const$ and $\mathcal{P}=const$ which introduces two constraints for the optimization problem. The solution then represents a grand canonical Gibbs distribution (see e.g. \cite{rckg00}):
\begin{equation}
\label{Gibbs}
P(\{a_n,a^*_n,b_n,b_n^*\}) =\mathcal{Z}^{-1}\,\exp\left[-\beta(\mathcal{H}-\mu \,\mathcal{P})\right]
\end{equation}
where the Lagrange multipliers $\beta$ and $\mu$ play the roles of the ``inverse temperature'' and ``chemical potential'' respectively while the normalizing factor $\mathcal{Z}$ has the familiar meaning of the partition function.

The assumption of field thermalization implies that instead of averaging the dynamics of the system (\ref{vector-dnlse}) over a long interval of $z$ one can compute the same averages via the equilibrium Gibbs measure (\ref{Gibbs}). However from the point of view of the nonlinear optical waveguides it is impractical to use averaging over large distances since it requires optical waveguides that are way too long. Instead the averaging can be understood as averaging over \textit{disordered initial conditions} that can be be experimentally controlled \cite{slbsm09,sps11}. This brings about the notion of the thermalization distance \cite{sps11} $z_{th}$ after which the information about the initial state of the system is forgotten and the averaging over the initial conditions is equivalent to Gibbs averaging. In our treatment we will assume (unless otherwise specified) that the initial amplitudes for both TE and TM components are constant $|a_n|=a$, $|b_n|=b$ while the phases are uncorrelated uniformly distributed random variables. Such assumption is not necessary but it simplifies the calculation of the ensemble averages of the initial Hamiltonian and power.

The knowledge of the partition function $\mathcal{Z}= \int \prod_n d \,a_n da_n^*\,db_n\,db_n^* \exp\left[-\beta(\mathcal{H}-\mu \,\mathcal{P})\right]$
allows one to calculate an average energy per waveguide $h=\mathcal{H}/N$ and the average intensity per waveguide $p=\mathcal{P}/N$. On the other hand such averages must correspond
to their initial values $h_0$, $p_0$ (averaged over the disordered initial conditions) since for each realization of the disorder these are conserved integrals of motion. Indeed, from (\ref{Gibbs}) it follows that
\begin{equation}
h_0=-\frac{1}{N}\,\frac{\partial \ln \mathcal{Z}}{\partial \beta} + \mu \,p_0, \quad p_0=\frac{1}{\beta\,N} \,\frac{\partial \ln \mathcal{Z}}{\partial \mu}
\label{h-p}
\end{equation}
For each set of phase-averaged initial conditions (i.e. given pair $(h_0,p_0)$) the above two transcendental equations yield the effective inverse temperature $\beta$ and the chemical potential, $\mu$. As in the scalar case \cite{rckg00} the necessary condition for the Gibbs distribution (\ref{Gibbs}) to be normalizable is the positiveness of the temperature, $\beta>0$. Thus the curve $\beta=0$ in the $(h_0,p_0)$ diagram represents a natural boundary between the conventional thermalization region ($\beta>0$) and the area where $\beta<0$ and the energy is localized in a form of discrete breathers that have been observed experimentally \cite{mhcssma03}.

It is convenient to introduce complex amplitudes and phases $a_n=\sqrt{A_n}\,\exp(i\,\phi_{1,n})$, $b_n=\sqrt{B_n}\,\exp(i\,\phi_{2,n})$ with $\delta_n\equiv \phi_{2,n}-\phi_{1,n}$ being the phase difference between the two orthogonal components. The case $\delta_n = m\pi$ ($m$-integer) corresponds to the linear polarization in $n$-the waveguide, $\delta_n=\pi/2$, $A_n=B_n$ describe circular polarization etc. This notation corresponds to the \textit{Jones description} of polarization (see e.g. \cite{Yariv}). To calculate the partition function $\mathcal{Z}$ one needs to integrate the Gibbs exponential in (\ref{Gibbs}) with the Hamiltonian (\ref{Hamiltonian-vect}). In the new variables the integral takes the form:
\begin{widetext}
\begin{equation}
\label{Z-vector}
\mathcal{Z} =\int \prod_n dA_n\, dB_n \,d\phi_{1,n} \,d\phi_{2,n} \exp\left[-\beta\left(k(A_n-B_n)+H_C(A_n,B_n,\phi_{1,n}-\phi_{1,n+1},\phi_{2,n}-\phi_{2,n+1})+H_\gamma(A_n,B_n,\delta_n)\right)\right]
\end{equation}
\end{widetext}
where we have collected all the nonlinear coupling terms as well as chemical potential in the nonlinear interaction part $\mathcal{H}_{\gamma}=\sum_n H_\gamma(A_n,B_n,\delta_n)$ with
\begin{equation}
\begin{split}
H_\gamma(A,B,\delta)&=\frac{\gamma}{2}\,\left[A^2+B^2+2\lambda_1 AB +2\lambda_2 AB \cos(2\delta) \right] \\
&-\mu\,(A+B)
\end{split}
\label{Hamiltonian-gamma}
\end{equation}
while the linear waveguide coupling is given by the Hamiltonian $H_C=\sum_n (\sqrt{A_n A_{n+1}}\cos(\phi_{1,n}-\phi_{1,n+1})+\sqrt{B_n B_{n+1}}\cos(\phi_{2,n}-\phi_{2,n+1}))$.

The Gibbs distribution (\ref{Gibbs}) is normalizable when the partition function is finite. A close inspection of Eq.(\ref{Z-vector}) reveals that in order to achieve this not only the temperature must be positive $\beta>0$, but additionally the inequality
\begin{equation}
\label{thermalization-ineq}
|\lambda_1-\lambda_2| < 1
\end{equation}
must hold. Already we can see a departure from the scalar case \cite{rckg00} where the Gibbs distribution is always normalizable as long as the temperature remains positive. For the isotropic case $\lambda_2=1-\lambda_1$ this corresponds to possible thermalization for $0 < \lambda_1 < 1$ and for the anisotropic case $\lambda_2=\lambda_1/2$ this implies $-2 < \lambda_1 <2$.  The borderline case $|\lambda_1-\lambda_2|=1$ (which includes the XPM without FWM $\lambda_1=1$, $\lambda_2=0$) generally requires special treatment and we will not consider it here. The non-normalizable property of the Gibbs distribution indicates the emergence of the localized structures, i.e the genuine equilibrium state now consist of high-amplitude discrete breather (or several breathers if the system has not yet quite reached the equilibrium) interacting weakly with a small quasilinear background thermalized at infinite temperature, $\beta=0$  \cite{jtz00,r08,r09}. So here the statistical mechanics  provides us with a clue as to the regions in the parameter space where the localized  structures can be observed in principle \cite{rckg00,jr04}.

In this paper we will only study the thermalization regime where the inequality (\ref{thermalization-ineq}) is fulfilled and the temperature is positive so that Gibbs distribution (\ref{Gibbs}) is always normalizable. We leave the analysis of the localized structures for future studies although we will comment on these in the following section. Since it is impossible to evaluate the partition function $\mathcal{Z}$ in the closed form we will study 3 different limiting regimes which for the scalar case were already analysed in Refs. \cite{rckg00,jr04,slbsm09}. These regimes are : i) low temperature limit $\beta \to \infty$ ii) high temperature limit $\beta \to +0$ (which also serves as a borderline for emergence of localized structures) and iii) the anticontinuum (high intensity) regime when the effective nonlinearity parameter $\Gamma$ defined below in section \ref{sec:nonl} is large, $\Gamma \gg 1$. It turns out that doubling the amount of degrees of freedom as compared to the scalar case has a significant impact on the statistical properties of system, (\ref{vector-dnlse}). We start our analysis with the first two regimes: $\beta \to \infty$ and $\beta \to 0$.

\section{The temperature boundaries and the thermalization region}
\label{sec:boundaries}
\begin{figure}
\includegraphics[width=86mm]{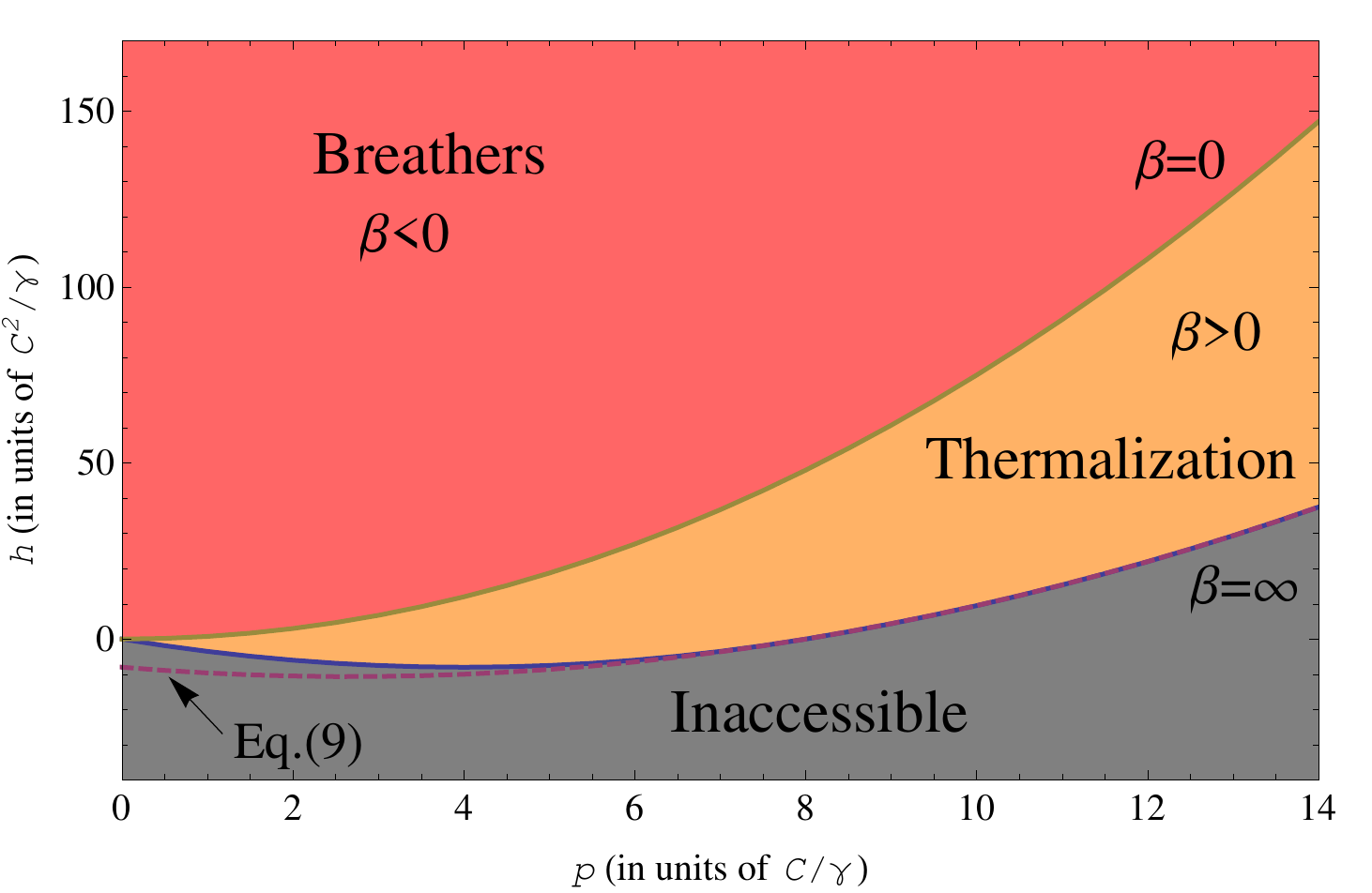}
\caption{\label{fig:h-p} (Color online) The different regions in $(h,p)$ parameter space. The ratio $k/C=2$ was assumed.}
\end{figure}

The limit of $\beta \to \infty$ corresponds to the configuration that minimizes the Hamiltonian (\ref{Hamiltonian-vect}) subject to the given conserved pulse intensity per waveguide $p$.  If we restrict our search to a solution with constant amplitudes, phase shifts and locked state of polarization (i.e. fixed $\delta$) the minimal configuration is achieved by the following field distribution: $a_n=\sqrt{A}\,\exp(i\,\pi\,n)$, $b_n=\sqrt{B}\,\exp(i\pi\,n+i\pi/2)$ where the amplitudes $A$ and $B$ as well as chemical potential (i.e. the corresponding Lagrange multiplier), $\mu$, are given by:
\begin{equation}
\label{min-amplitudes}
\begin{split}
A&=\frac{p}{2}-\frac{k}{\gamma}\frac{1}{1-\lambda_1+\lambda_2}, \quad B=\frac{p}{2}+\frac{k}{\gamma}\frac{1}{1-\lambda_1+\lambda_2} \\
\mu & = \frac{1}{2}\,p\,\gamma\,(1+\lambda_1-\lambda_2).
\end{split}
\end{equation}
This solution exists only for not too low intensities, i.e. for $p> (2k/\gamma)(1-\lambda_1+\lambda_2)^{-1}$ and provides a low bound for the energy per waveguide:
\begin{equation}
h_{min}=-2pC+\frac{(1+\lambda_1-\lambda_2)\gamma}{4}\,p^2-\frac{k^2}{\gamma(1-\lambda_1+\lambda_2)}.
\label{hmin}
\end{equation}
In the limit of zero birefringence, XPM and FWM, $\lambda_1=\lambda_2=k=0$ we get the energy as a sum of energies of two identical scalar DNLSEs with the average intensity per waveguide $a/2$ - which corresponds to the result of Ref.\cite{rckg00}. In Fig. \ref{fig:h-p} (which is an analogue of Fig.1 of Ref.\cite{rckg00}) we plot a phase diagram in the $(h,p)$  space for a specific case of 16 waveguides with $\lambda_1=1$, $\lambda_2=0.5$. Both $h$ and $p$ have been rescaled to the dimensionless multiples of $C^2/\gamma$ and $C/\gamma$ respectively. One can see that Eq.(\ref{hmin}) provides an excellent low boundary approximation even below the critical intensity for which the solution of (\ref{min-amplitudes}) exists (which is $p=8$ for the chosen parameters).
Let us now turn to the opposite case of high temperatures $\beta \to 0$ assuming that the product $\beta \mu <0$ remains finite. We will use the method similar to that of Ref.\cite{jr04} applied earlier to the scalar case. In this limit one can neglect the linear coupling energy $H_C$ in the exponent of  (\ref{Z-vector}) together with the birefringence contribution $k (A_n-B_n)$. The partition function is then given by $\mathcal{Z}=(4\pi^2 y(\beta,\mu))^N$ with
\[
\begin{split}
y(\beta,\mu)= &\intop_0^\infty dA \,\intop_0^\infty dB\, \mathrm{I}_0 \left[\beta \,\gamma\,\lambda_2 \,A\,B\right] \\
\times & \exp\left[-\frac{\beta\gamma}{2}\,\left(A^2+B^2+2\lambda_1 \, A \, B \right) -\beta\,|\mu| \,(A+B)\right].
\end{split}
\]
Next we assume that $\lim_{\beta\to 0}(\beta\gamma)=0$ and write approximately
\begin{eqnarray*}
y(\beta,\mu) &= &\intop_0^\infty dA \intop_0^\infty dB \exp(-\beta|\mu|(A+B)) \\
& \times & \left(1-\frac{\gamma\,\beta}{2}\,\left(A^2+B^2+2AB\lambda_1\right)\right) + O[(\beta\gamma)^2],
\end{eqnarray*}
where we have used the fact that the product $\beta|\mu|$ is fixed while $\beta\gamma$ tends to zero. The answer is
\[
y(\beta,\mu) = \frac{1}{(\beta\mu)^2}-\frac{\beta\gamma(2+\lambda_1)}{(\beta \mu)^4} + \ldots
\]
Finally from Eqs.(\ref{h-p}) we obtain in the leading approximation:
\begin{equation}
\label{beta-zero}
\lim_{\beta \to 0} (\beta \mu) =-2/p, \quad h_{max}=\frac{\gamma(2+\lambda_1)}{4} \,p^2.
\end{equation}
The parabola $h_{max}(p)$ provides the upper boundary $\beta=0$ for the thermalization region in the plane of parameters $(h,p)$ - see Fig.\ref{fig:h-p}. Interestingly enough it does not depend on the 4-wave mixing constant $\lambda_2$.

\section{Nonlinearity dominated regime}
\label{sec:nonl}
In the following chapters we will adopt normalized units where the propagation distance is measured in units of coupling length $z \to C\,z$. We will also assume that the intensities of both TE and TM components are $A_n(0)=A_0$, $B_n(0)=B_0$. We can also normalize the intensities of both components by the half of the initial intensity $p_0/2$ so that $A_0+B_0=2$. In the new dimensionless units one must simply substitute $k/C$ for $k$, $1$ for $C$ in the original coupled equations (\ref{vector-dnlse}) and instead of the nonlinear coefficient $\gamma$ one now has a dimensionless nonlinearity parameter $\Gamma = \gamma p_0/(2C)$. Since in most common nonlinear materials and waveguide geometries the ratio $k/C$ is in the order of unity the parameter $\Gamma$ indeed gauges the relative strength of nonlinearity \cite{slbsm09,sps11}.

In this section we will be interested in the highly nonlinear regime, $\Gamma \gg 1$. The motivation for this is twofold. Firstly, this regime permits almost full analytical treatment which is always helpful when studying the general properties of any nonlinear system and secondly, in Ref.\cite{slbsm09} it was shown that in the scalar case this limit corresponds to the reciprocal temperature $\beta$ that does not depend on the value of parameter $\Gamma$ and is a universal constant (in dimensionless units). The universality of the temperature in turn gives rise to a universal shape of the field correlation function. Therefore it is interesting to see how this result changes in the vector case. One should expect that the dynamics and statistics in the vector case are much richer due to the doubled number of interacting degrees of freedom. Here we will show that this is indeed the case.

Our main objects of interest will be the statistics of the intensity and polarization state of each waveguide as well as the distribution of the phase differences (i.e. phase gradients in the continuous limit) between the same components (e.g. TE) of the adjacent waveguides together with the field correlation functions. It turns out that in the strongly nonlinear regime the statistics of the phase gradient is decoupled from those of the intensity and polarization. The latter are most conveniently described by using a popular alternative to the Jones description of polarization, namely by introducing the four Stokes parameters $\{S_i\}_{i=0}^3$ \cite{contvectdnls,aa,Yariv}. They are related to the Jones parameters via:
\begin{equation}
\label{Stokes}
\begin{split}
S_0^n&=A_n+B_n\\
S_1^n&=A_n-B_n\\
S_2^n&=2\,\sqrt{A_n\,B_n}\cos(\delta_n)\\
S_3^n&=-2\,\sqrt{A_n\,B_n}\sin(\delta_n)\\
\end{split}
\quad \sum_{i=1}^3 (S_i^n)^2=(S_0^n)^2.
\end{equation}
The vector $\vec{S}^n=(S_1^n,S_2^n,S_3^n)$ is called a Stokes vector on a Poincar\'{e} sphere of radius $S_0^n$. The components of the Stokes vector for each waveguide are related to the polarisation state of the waveguide while its magnitude provides the total intensity carried by both field components. For example,  linear polarization corresponds to the equatorial plane $S_3^n=0$ while the circular clockwise and anticlockwise polarizations correspond to the north and south poles $\vec{S}^n=(0,0,\pm S_0^n)$. In what follows we will use both Jones and Stokes descriptions whichever is is more convenient.

\subsection{The statistical properties of the field in the anticontinuum limit ($k=C=0$)}
\label{sec:anticont}

We will start our analysis with the anticontinuum limit when one can completely neglect both linear coupling and birefringence in Eqs.(\ref{vector-dnlse}). As will be seen later this corresponds to the initial stages of evolution of field distribution towards the final state of equilibrium.
The absence of linear waveguide coupling makes field dynamics in each waveguide independent from the rest. Therefore instead of considering the system of coupled field equations (\ref{vector-dnlse}) one can analyze the dynamics in a single waveguide. In particular the first three Stokes parameters obey the following equations (the waveguide-index has been omitted for convenience):
\begin{equation}
\label{Stokes-dyn}
\begin{split}
\frac{d S_1}{d z} & = 2\,\Gamma\,\lambda_2\,S_2\,S_3 \\
\frac{d S_2}{d z} &=- \Gamma(1+\lambda_2-\lambda_1)\,S_1\,S_3 \\
\frac{d S_3}{d z} & =\Gamma (1-\lambda_2-\lambda_1)\,S_1\,S_2
\end{split}
\end{equation}
This system is completely integrable and its various special cases and generalizations (like e.g. the inclusion of the birefringence terms $\sim k$) have been extensively studied in literature both with connection to the nonlinear polarization dynamics \cite{couplers} in general and the dynamics of the polarization-locked vector solitons \cite{contvectdnls,aa} in particular. The complete integrability of system (\ref{Stokes-dyn}) is due to the existence of two integrals of motion of which the first is just the intensity, i.e. the zeroth Stokes parameter, $S_0$, and the second, $R$, is related to the Hamiltonian of the original system \cite{couplers}:
\begin{equation}
\begin{split}
S_0^2&=S_1^2+S_2^2+S_3^2, \\
R& = \left\{\begin{array}{cc} (3\lambda_1/2-1)S_2^2-(1-\lambda_1/2)\,S_3^2, & \mbox{anisotropic} \\
& \\
S_3, & \mbox{isotropic}\end{array} \right.
\end{split}
\label{Stokes-integrals}
\end{equation}
In other words the dynamics of the system takes place on the intersection of the Poincar\'{e} sphere of radius $S_0$ and a hyperbolic (or elliptic) cylinder in the anisotropic case or a plane in the isotropic case given by the equation $R=const$. One can obtain an autonomous equation for $S_1$ which has the form of a Duffing oscillator equation in the anisotropic case \cite{couplers} and harmonic oscillator equation in the isotropic case \cite{aa} (see also Appendix \ref{sec:appendix}). The other two Stokes parameters are recovered from Eqs.(\ref{Stokes-dyn}) while the phase of the TE component can be determined by simple integration of the original field equations. Since the system in the anticontinuum limit is completely integrable it does not thermalize, i.e. formula (\ref{Gibbs}) is inapplicable. Instead one must use the exact solution for the Stokes vector $\vec{S}(z)$ and average it directly over the initial conditions.

\begin{figure*}[tb]
\includegraphics[width=178mm]{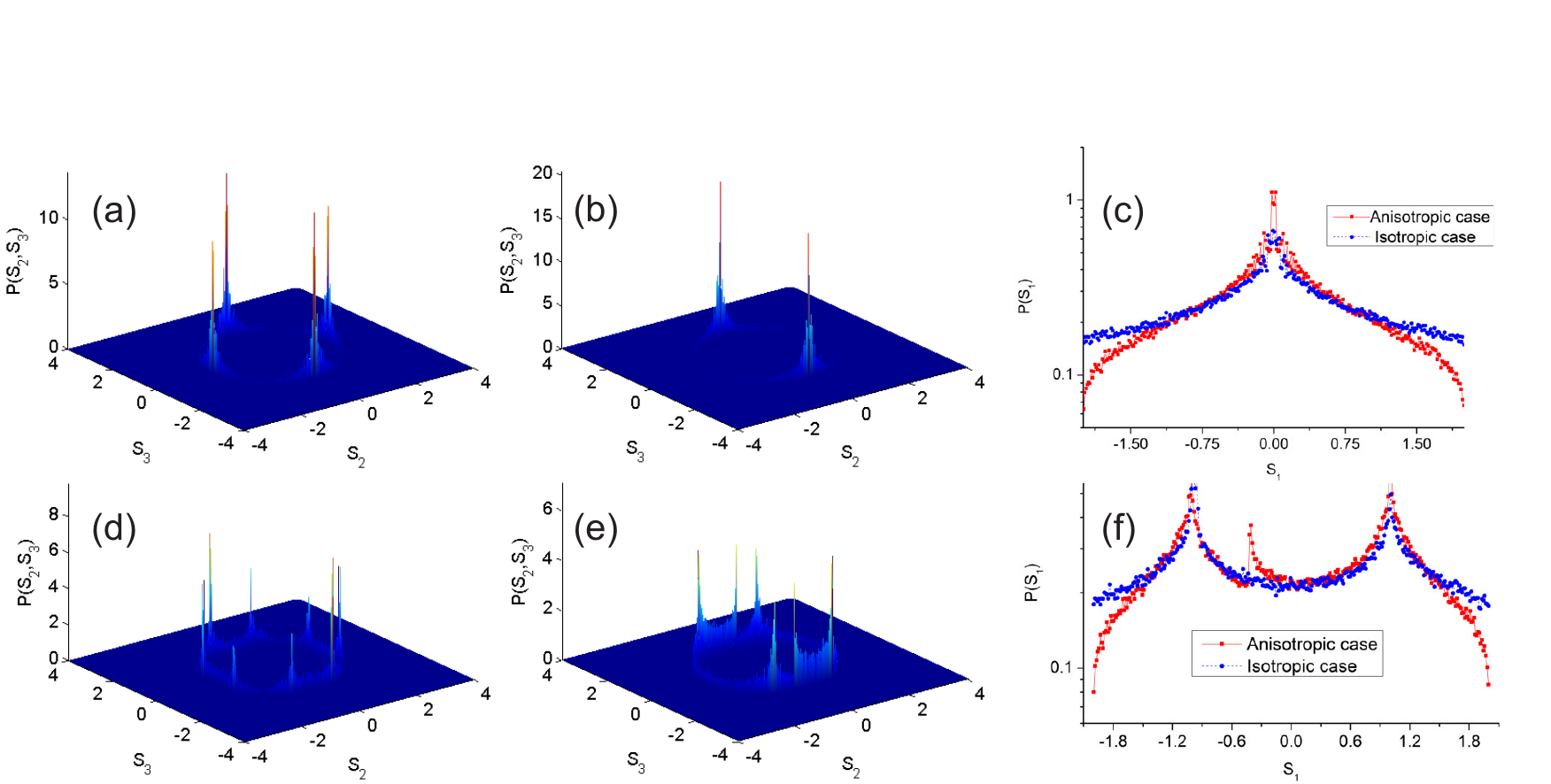}
\caption{\label{fig:anticont} (Color online) Marginal probability density functions for different Stokes parameters evaluated for the nonlinearity parameter $\Gamma=50$ at distance $z=10$ (in units of coupling lengths). Top row corresponds to the initial equipartition of intensity between the components $A_0=B_0=1$ while the bottom row shows the case when $A_0=1.5$ and $B_0=0.5$.  (a) and(d) - correspond to anisotropic case with $\lambda_1=1$, $\lambda_2=1/2$, (b) and (e) assume isotropic case with $\lambda_1=2/3$, $\lambda_2=1/3$. Marginal PDF $P(S_1)$ is plotted in (c) and (f).}
\end{figure*}
As mentioned earlier we will assume that all waveguides initially have the same set of intensities, $A_0$, $B_0$, $S_0=A_0+B_0=2$ and random, independent phases. From the definition of the Stokes parameters (\ref{Stokes}) it follows that that initial value $S_1(0)=A_0-B_0$ is fixed while the points $(S_2(0),S_3(0))$ are uniformly distributed on a circle of radius $2\sqrt{A_0\,B_0}$. In Fig.\ref{fig:anticont} we show the evolution of the marginal probability density functions (PDFs) $P(S_2,S_3)$ and $P(S_1)$ for the anisotropic and anisotropic cases obtained by averaging the solution of system (\ref{Stokes-dyn}) over the initial phase distribution.  The parameters $\lambda_{1,2}$ were chosen to correspond to the AlGaAs compounds in the anisotropic case and fused silica for the isotropic case. One can see that the structure of the histograms is very sensitive to both the values of $\lambda$-coefficients and the symmetry of the initial condition. For example when the initial intensity is equally distributed between the components, i.e. $A_0=B_0=1$, $S_1(0)=0$ the marginal distribution $P(S_1,S_2)$ shown in Fig.\ref{fig:anticont}(a) is 4-modal. It is largely confined to the initial circle of radius $S_0=2$ and the four maxima are at the points $(0,\pm S_0)$, $(\pm S_0,0)$. At the same time the marginal PDF for the first Stokes parameter $S_1$ is sharply centered at zero (Fig.\ref{fig:anticont}(c)) i.e. for most initial realizations the symmetry between the components is preserved $A(z) \approx B(z)$. We have also checked this property for different runs with different distances, $z$ (not shown). This means that in the anticontinum limit for the anisotropic cubic crystal (like e.g. AlGaAs) each given waveguide evolves into either linear or circular state of polarization - despite that no such preference existed in the initial conditions (the distribution was uniform on the circle). In the isotropic case one can observe that although the distribution of $P(S_1)$ is still sharply peaked around zero (Fig.\ref{fig:anticont}(c)) the distribution of the two remaining Stokes parameters Fig.\ref{fig:anticont}(b) is only bi-modal (and not 4-modal as in the anisotropic case) with the two maxima at $(0,\pm S_0)$ corresponding to circular polarization only. The positions of the maxima in both cases can be easily explained by considering the two integrals of motion (\ref{Stokes-integrals}). As we have seen in both cases for most realizations one can neglect the value of $S_1$ and the PDF $P(S_1,S_2)$ therefore  remains confined to the circle of radius $S_0=2$ for all values of $z$. Next if one looks at the distribution of the second integral of motion in (\ref{Stokes-integrals}), namely $R$, then for the initial values $(S_2(0),S_3(0))$ uniformly distributed on a circle of radius 2 one can see that the distribution of $R$ is bimodal with the two maxima at $R_-=-4(1-\lambda_1+\lambda_2)$ and $R_+=4(\lambda_1+\lambda_2-1)$) in the anisotropic case and $R_\pm=\pm 2$ in the isotropic case. The intersection points of the pair of curves $R(S_2,S_3)=R_\pm$ with the circle in the $(S_2,S_3)$ plane are exactly the four observed maxima in the anisotropic case and two in the isotropic one.

In the case when initially the symmetry between the field components is broken (bottom row in Fig.\ref{fig:anticont}) the situation is much more complex owing to the multimodal features of the distribution of the parameter $S_1$ - see Fig.\ref{fig:anticont}(f). Each maximum of the $P(S_1)$ gives rise to at least 2 maxima in the distribution $P(S_2,S_3)$ producing complex crown-like shapes shown in Fig.\ref{fig:anticont}(d),(e). Each of the maxima now corresponds to a certain elliptically polarized state which varies from waveguide to waveguide. Note also that the positions of the maxima remain fixed with the propagation distance $z$ while their widths experience weak periodic oscillations (not shown). The full theoretical analysis of these multimodal distributions requires further study which is beyond the scope of this paper.

\subsection{The structure of the final thermal state. The regime of the universal temperature}
\label{sec:universal}
We have seen above that in the anticontinuum limit when the birefringence and the linear waveguide coupling can be neglected completely the system is integrable and instead of reaching thermal equilibrium it experiences periodic oscillations that can be averaged directly over the initial conditions to produce multimodal distribution for the established state of polarization. The question then arises as to what will happen if both the birefringence and the coupling are taken into account. One can expect that at the initial stages of evolution when the distance is less than the thermal length, $z_{th}$, the evolution of the system is close to anticontinuum limit and the state of polarization follows the statistics derived above for the anticontinuum limit. For $z \gtrsim z_{th}$ however, the nonintegrability of the system becomes essential. The linear coupling between the waveguides leads to energy and power exchange between the waveguides so that eventually most of the initial peaks in the marginal PDF $P(S_2,S_3)$ are destroyed and final thermal distribution sets in. This final distribution is characterized by the Gibbs statistics (\ref{Gibbs}) with the partition function given by (\ref{Z-vector}).

We wish to calculate the partition function and the probability distributions in this final state still assuming strong nonlinearity $\Gamma \gg 1$ but now taking into account the linear coupling and birefringence as well. Let us first assume that the initial conditions are such that the system thermalizes into a state with finite temperature, i.e. $\beta \to const$ as $\Gamma \to \infty$. We will shortly see that this is only possible for a very special choice of the initial conditions. Then the dominating part of the exponent in (\ref{Z-vector})  is the nonlinear interaction term $\mathcal{H}_\gamma$. Other terms (including the phase-dependent coupling energy) are relatively slow functions of the amplitudes and do not therefore contribute to the field intensity PDF and the partition function. However it is important to retain the chemical potential term since as we shall see below $\mu \sim \Gamma$ always. When calculating the partition function (\ref{Z-vector}) one can resort to the saddle point approximation in $\gamma=\Gamma$ which implies that at large values of $\Gamma$ the main contribution to the integral comes from absolute minimum of $H_\gamma(A_n,B_n,\delta_n)$.  This minimum is achieved for the values
\begin{equation}
\label{min-thermal}
A_*=B_*=\frac{\mu}{\Gamma(1+\lambda_1-\lambda_2)},\quad  \delta_*=\pm\pi/2.
\end{equation}
In other words the absolute minimum of the interaction Hamiltonian subject to given total power is achieved by a circularly polarized state - the most symmetric of all. This minimum is also degenerate - for a system of $N$ waveguides there are $2^N$ possible choices of polarization orientation. This degeneracy is drastically reduced however if one takes into account the coupling term $H_c$. This term favors the configuration where the phase difference $\delta_n$ is uniform across the waveguides which leaves only two possibilities: either all fields are clockwise polarized or they are all anticlockwise polarized. These states also corresponds to the two maxima of the distribution function $P(A,B,\delta)$ (or $P(S_1,S_2,S_3)$ in Stokes parameters) one for each direction of rotation. As previously we assume here that $|\lambda_1-\lambda_2|<1$ which ensures the convergence of the integrals for positive temperature. The minimal value of the interaction energy (at fixed power) is given by
\[
H_\gamma(A_*,B_*,\delta_*)=-\frac{\mu^2}{\Gamma(1+\lambda_1-\lambda_2)}.
\]
If we assume uniform initial amplitudes $A_0$ and $B_0$ ($A_0+B_0=2$) and neglect the terms that are not proportional to $\Gamma$ in the limit $\Gamma \gg 1$ the average energy per waveguide is given by:
\begin{equation}
h_0=\Gamma\left[\frac{A_0^2+B_0^2}{2} + A_0B_0(\lambda_1+\lambda_2\langle \cos(2\delta_0)\rangle)\right].
\label{def-f}
\end{equation}
The energy is always bound from below by $h_{min}=\Gamma(1+\lambda_1-\lambda_2)>0$. In the spirit of saddle point approximation we can now plug $A_n=A_*$, $B_n=B_*$, $\delta_n=\delta_*$ from Eq.(\ref{min-thermal}) into linear coupling pre-factor $\exp (-\beta H_{C})$ when calculating the partition function. The integration over the phases $\phi_{1,n}$ of the TE component is reduced then to calculation of the partition function of the 1D calssical XY model (see \cite{rckg00} or \cite{m84} for details). In the thermodynamic limit ($N\to \infty$) the contribution of this linear coupling term to the logarithm of the partition function is $- N \ln I_0(4\beta\mu/(\Gamma(1+\lambda_1-\lambda_2)))$. The remaining Gaussian integration over the fluctuations around the minimum of $H_\gamma$ Eq.(\ref{min-thermal}) is trivial. Inserting the resulting expression for the partition function $\mathcal{Z}$ into the system of equations (\ref{h-p}) after some simple algebra we obtain the following transcendental equation for the inverse temperature $\beta$:
\begin{equation}
h_0-h_{min} =\frac{3}{2\beta}-\frac{4\mathrm{I}_1(4\beta)}{\mathrm{I}_0(4\beta)}, \quad \mu=\Gamma(1+\lambda_1-\lambda_2).
\label{beta-saddle}
\end{equation}
with $h_0$ given by Eq.(\ref{def-f}).
For the consistency of our approximation we must demand that the inverse temperature $\beta$ remains finite as $\Gamma$ goes to infinity (as is the case in the scalar DNLSE \cite{slbsm09}). But this is achieved only when $h_0=h_{min}$ so that the l.h.s. vanishes which means that the initial input must be circularly polarised: $A_0=B_0=1$, $\delta_0=\pm \pi/2$.  In other words, in order to obtain a universal (i.e. $\Gamma$-independent) temperature constant, similar to the scalar case, the initial state must necessarily be the one that minimizes the nonlinear part of the Hamiltonian, $H_\gamma$ subject to the intensity constraint $A_0+B_0=2$. The system then becomes effectively scalar and thermalization occurs only in the distribution of the TE phase differences $\theta_n=\phi_{1,n}-\phi_{1,n+1}$ while the field in each waveguide remains locked in its original clock- or anti-clockwise circular state of polarization. As in Ref.\cite{slbsm09} the Gibbs distribution can be approximately factorized so that the intensity of each waveguide $P_n=A_n+B_n=S_0^n$ has a narrow distribution close to Gaussian centered around the conserved mean value of $2$. 
\begin{figure}[tb]
\includegraphics[width=86mm]{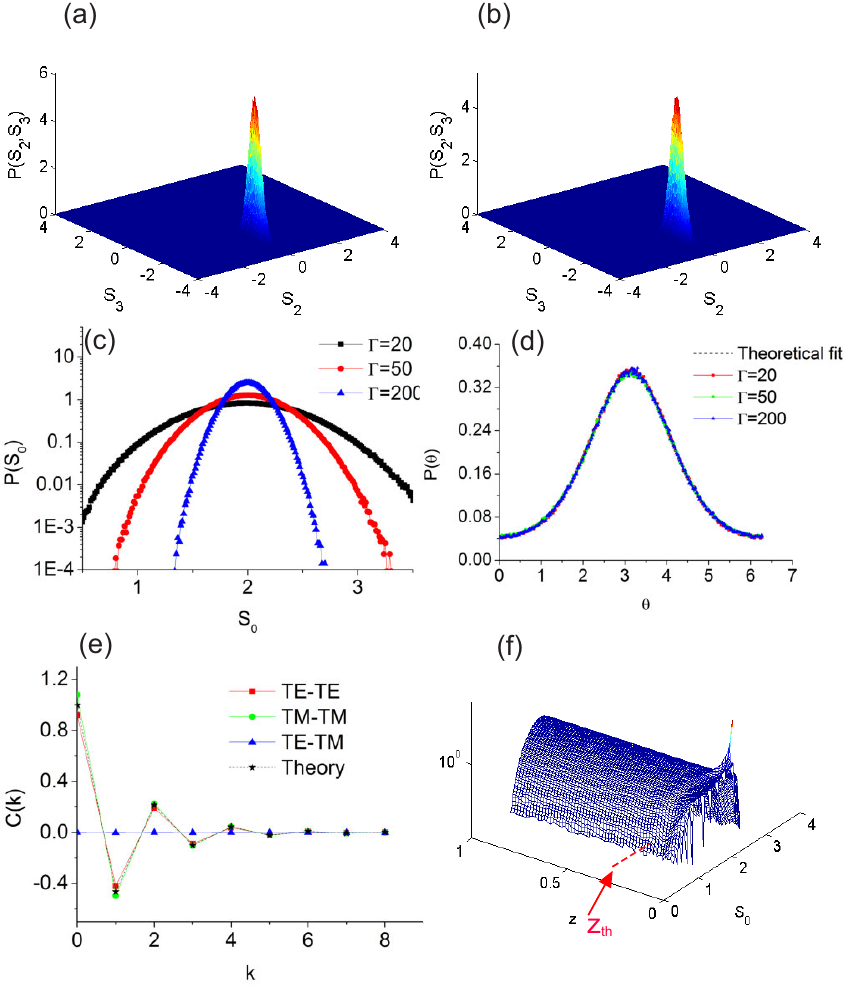}
\caption{\label{fig:universal} (Color online) The statistics of the final thermal state in the limit of large nonlinearity when the initial field in \textit{each} waveguide is circular clockwise polarized. (a),(b) The marginal distribution of Stokes parameters $P(S_2,S_3)$ for anisotropic and isotropic case respectively. (c), (d) The statistics of the intensity and phase difference (of the TE component) respectively for different values of the nonlinearity parameter. (e) Cross and self field correlation functions for the TE and TM components. (f) The evolution of the intensity PDF with distance. The thermalization length $z_{th}$ is determined as a point where the distribution stabilizes.}
\end{figure}
The results of the numerical simulations of universal correlations in the vector case of $N=64$ waveguides are presented in Fig. \ref{fig:universal}. As prescribed, we have started from a symmetric initial configurations $A_0=B_0=1$ and uniform, uncorrelated phase distribution for the TE components. To ensure constant circular polarization locking the phases of the TM component were obtained by those of the TE by adding $\pi/2$ (corresponding to the clockwise field rotation). In all the figures (unless otherwise specified) the values of the parameters are: $k=2$, $\Gamma=50$ and the propagation length is $z=10$ (all in the normalized units). In Fig.\ref{fig:universal} (a),(b) we show numerical simulations of the marginal PDF $P(S_2,S_3)$ for the anisotropic ($\lambda_1=1$,$\lambda_2=1/2$) and isotropic ($\lambda_1=2/3$, $\lambda_2=1/3$) cases respectively. One can see that both distributions have narrow maxima at the point $(0,-2\sqrt{A_* B_*})=(0,-2)$ corresponding to clockwise circular polarization (as in the initial state). We have found very little qualitative difference between the anisotropic and isotropic cases so Figs.\ref{fig:universal}(c)-(f) feature only the former. In \ref{fig:universal}(c) we can see that the intensity PDF becomes more and more narrowly centred around the average value of 2 and the distribution is close to Gaussian with the variance proportional to $(\beta \Gamma)^{-1}$. As for the PDF for the TE phase difference, $P(\theta)$, the theoretical prediction $P(\theta)=(2\pi I_0(4\beta))^{-1} \, \exp(-4\beta \cos \theta)$ is clearly corroborated by the numerics shown in \ref{fig:universal}(d): one can observe a perfect data collapse for different values of $\Gamma$. Similar to universal correlations in the scalar case, one can obtain universal filed correlations in the vector case. In \ref{fig:universal}(e) we plot the field correlation functions functions defined as $C_{qq'}(k)=\langle \mathrm{Re} [q_n q'^{*}_{n+k}] \rangle $ where $q$ ,$q'$ denote different components of the field, TE or TM, and the average is taken both over the waveguide position, $n$, and the initial conditions. Since in the thermal equilibrium the TE and TM components are always $\pi/2$ out of phase the cross-field correlation $C_{ab}(k)$ is always zero while the self-field correlation functions have identical universal shape:
\begin{equation}
\label{field-corr}
C_{aa}(k)=C_{bb}(k)=\eta^k, \quad \eta =\langle \cos \theta \rangle =-\frac{\mathrm{I}_1(4\beta)}{\mathrm{I}_0(4\beta)}.
\end{equation}
This is illustrated in Fig.\ref{fig:universal}e, where all three components are plotted together with the theoretical fit. Finally, Fig.\ref{fig:universal}(f) illustrates how quickly the final thermal distribution is achieved. The initial delta-peak in intensity distribution quickly broadens and settles to a stationary, almost Gaussian distribution of the type shown in \ref{fig:universal}(c) already at the point $z \approx 0.3$ (in units of the coupling length) that naturally serves as the thermalization length $z_{th}$. One can see that thermalization here occurs over just a fraction of the coupling length, i.e. on the scale of $0.15$ mm for a typical AlGaAs waveguide system. Of course one can come up with a more rigorous quantitative definitions of $z_{th}$, by looking e.g. at the saturation of the temperature defined from the simulated data via equipartition theorem \cite{sps11} but for the purposes of this paper we can restrict ourselves to the crude estimate above. All other distributions also stabilize at the same length.

As for the value of the universal temperature constant, $\beta$, one might think that it can be obtained from Eq.(\ref{beta-saddle}) by putting the l.h.s. to zero. Unfortunately this is not so since the resulting value of $\beta$ differs from the one observed in the numerics ($\beta \approx 0.26$) by the factor of almost 2. The reason for such a discrepancy is that the vanishing l.h.s. of Eq.(\ref{beta-saddle}) generally represents the main terms of approximation ($\sim \Gamma$). As these terms have now cancelled each other the universal temperature $\beta \sim 1$ is given by the balance of the next terms in the expansion that are of the order of unity. But it turns out that the neglected terms in the saddle point approximation of the partition function being of the order  $O(1/\Gamma)$ have nevertheless the contribution to its logarithmic derivative that are in the order of 1, which affects the value of the reciprocal temperature $\beta$.  Therefore in principle one has to consider the next terms in the saddle point approximation of (\ref{Z-vector}). However we have opted for a different approach, namely, we have evaluated the logarithmic derivatives of the integral of $\exp(-\beta H_\gamma)$ numerically and the obtained solution of system (\ref{h-p}) provided us a value of $\beta$ which fitted well the dependencies shown \ref{fig:universal}(d)-(e) (and also the intensity PDFs in \ref{fig:universal}(c)). From a few separate runs (not shown) it appears also that not only the found universal value $\beta \approx 0.26$ is independent of $\Gamma$ (when $\Gamma \gg 1$) but it is apparently independent of the material parameters $\lambda_{1,2}$ as well. This fact however requires further thorough verification.

\subsection{The structure of the final thermal state. General case}
\label{sec:thermal-general}
In the above section we have shown that if one ensures that the initial field components are locked in the same circular state of polarization for each waveguide, the reciprocal temperature is a universal numerical constant $\beta$, independent on the nonlinearity parameter $\Gamma$ which implied universal shape of the intensity and phase PDFs as well as the field correlation functions. Let us now turn to the more general case when the initial field does not have any preferred state of polarization, i.e. although both TE and TM components have constant intensities $A_0$, $B_0$ ($A_0+B_0=2$) their phases are always independent and uniformly distributed. It turns out that although the final thermal state is still given by the constrained minimum of the nonlinear energy corresponding to the circular polarization given by (\ref{min-thermal}), the value of the temperature $\beta$ is no longer a universal constant but depends on the nonlinearity. Also there is a drastic increase in the thermalization length, sometimes by up to 3 orders of magnitude. However we shall see below that in this general case the ``universality'' is not lost altogether. Rather it is the product $\beta \Gamma$ and not the temperature itself that is universal so instead of e.g. universal filed correlations as in the scalar case \cite{slbsm09} one has \textit{universal intensity distribution} $P(S_0)$ that is not affected by the value of the nonlinearity parameter $\Gamma$. The width of this distribution (i.e. the variance of the intensity fluctuations) is of the order of $\beta \Gamma \sim 1$. It turns out, however, that this new constant $\beta \Gamma$ is less ``universal'' than the temperature in the above-considered case of initial circular polarization inasmuch as it exhibits strong dependence on parameters $\lambda_{1,2}$.

We start by noticing that from Eq.(\ref{beta-saddle}) it follows that if the initial fields are not circularly polarised, i.e. $h_0-h_{min} \sim \Gamma \neq 0$ the reciprocal temperature $\beta$ must necessarily be of the order $\Gamma^{-1} \ll 1$ to balance the large terms in the l.h.s. Of course (\ref{beta-saddle}) was obtained in the saddle point approximation which strictly speaking no longer applies when the product $\beta \Gamma$ is in the order of unity.
But as we shall presently see the relation $\beta \,\Gamma \sim 1$ also follows from the exact scaling dependence of the partition function valid beyond the saddle point approximation. We will introduce the new notations for the re-scaled temperature $\tilde \beta = \beta \Gamma$, and chemical potential $\tilde \mu=\mu/ \Gamma$ that are now both in the order of 1. One can notice that in this regime since $\beta$ scales as $\Gamma^{-1}$ one can neglect the coupling energy, $H_C$ (as well as the birefringence term $k(A_n-B_n)$) in (\ref{Z-vector}) which simplifies the calculations somewhat. In fact this assumption is similar to the one used in section \ref{sec:boundaries} when we determined the infinite temperature boundary $\beta=0$ with the only difference being that now one cannot expand $\exp (\beta H_\gamma)$ in powers of the argument (since the latter is in the order of unity).

Regardless of the applicability of the saddle point approximation the numerical simulations demonstrate that a single waveguide probability density function factorises into the product of the PDF of the phase differences in the TE polarization, $P(\theta_n)$ and the PDF for the remaining three Jones parameters, $P(A_n,b_n,\delta_n)$, with:
\begin{equation}
\label{saddle-pdfs}
\begin{split}
P(\theta)=\frac{1}{2\pi\mathrm{I}_0(4\tilde \beta \tilde \mu/h_{min})}\, & \exp\left(-4\,\tilde \beta\,\tilde \mu \,\cos \theta /h_{min}\right) \\
& \\
P(A,B,\delta) & \propto e^{-\beta H_\gamma(A,B,\delta)}
\end{split}
\end{equation}
Note that since $h_{min}\sim \Gamma$ while $\tilde \beta \sim \tilde \mu \sim 1$ the phase PDF $P(\theta_n)$ is i) non-universal (i.e depends on the nonlinearity) and ii) approaches the uniform distribution as $\Gamma \to \infty$ (unlike the scalar case where it remains fixed, c.f. Fig.\ref{fig:universal}(d)). The intensity distribution is no longer narrow and has a finite width proportional to $\tilde \beta^{-1/2}$.

It is convenient at this stage to change variables from Jones to Stokes parameters  $(A,B,\delta) \to (S_1,S_2,S_3)$ according to the definition (\ref{Stokes}). The Jacobian of this transformation is equal to $(2S_0)^{-1}$ and the marginal PDF for the Stokes parameters has the form:
\begin{equation}
\label{pdf-Stokes}
\begin{split}
P(S_1,S_2,S_3) & \propto S_0^{-1} \exp\left[-\tilde \beta \left(\frac{1}{2}S_1^2 + \frac{1+\lambda_1}{4}\, (S_2^2+S_3^2) \right. \right. \\
& \left. \left. +\frac{\lambda_2}{4}\,(S_2^2-S_3^2)-\tilde \mu \, S_0\right)\right],
\end{split}
\end{equation}
and $S_0=\sqrt{S_1^2+S_2^2+S_3^2}$.
\begin{figure}[h!]
\includegraphics[width=86mm]{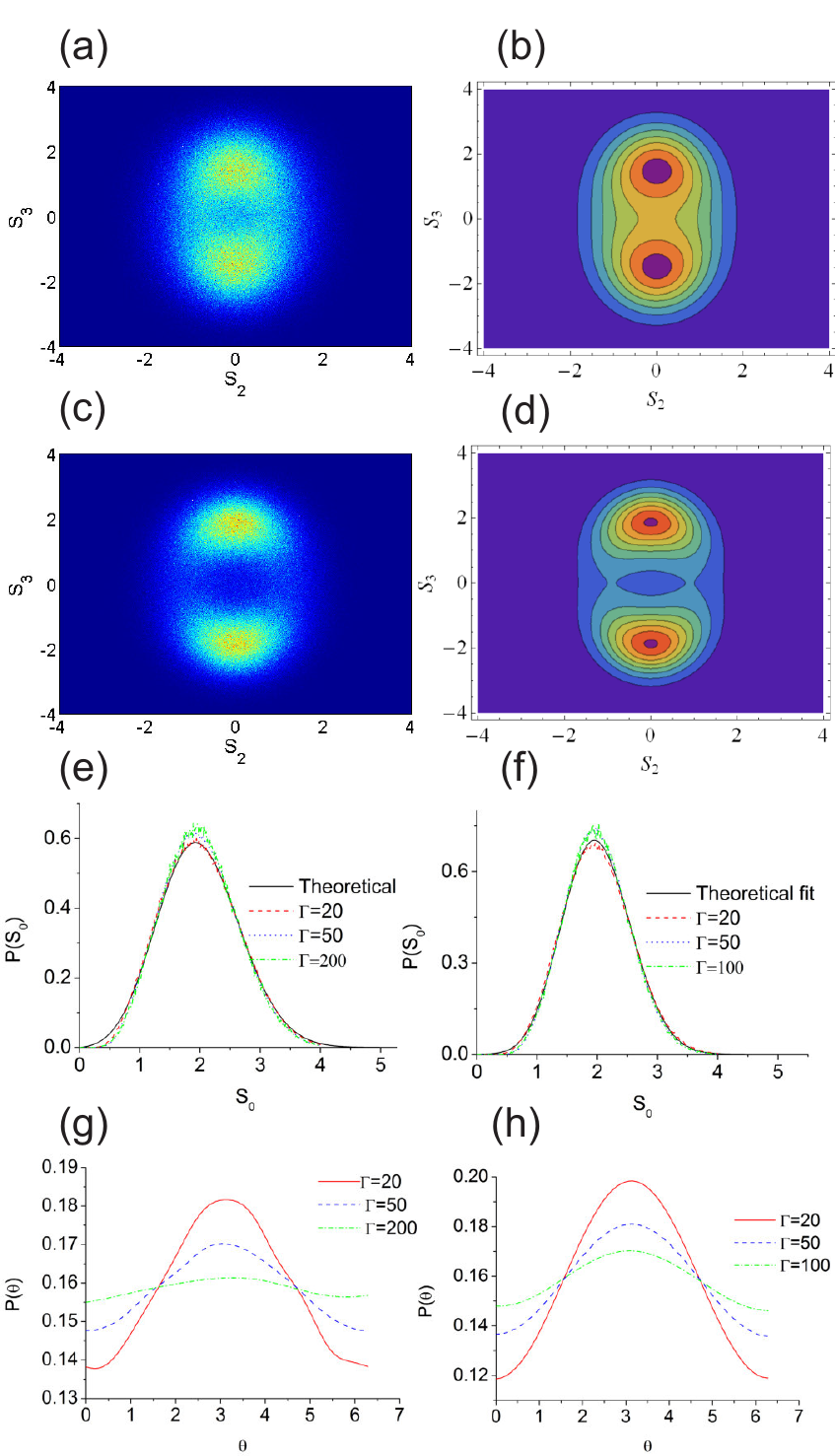}
\caption{\label{fig:general} (Color online) The statistics of the final thermal state in the limit of large nonlinearity and no initial phase correlation. (a)-(d) The marginal distribution of Stokes parameters $P(S_2,S_3)$ for anisotropic (a),(b) and isotropic (d,e) cases respectively. (a) and (c) show numerical simulations while (b) and (d) demonstrate theoretically calculated profiles. (e), (f) The statistics of the intensity
for anisotropic (e) and isotropic (f) cases for different values of nonlinearity parameter $\Gamma$. The phase difference distribution $P(\theta)$ for anisotropic (g) and isotropic (h) cases for different values of the nonlinearity parameter.}
\end{figure}

The marginal probability for the intensity as well as the overall normalization, the temperature and the chemical potential can be established by introducing the spherical coordinates on the Poincar\'{e} sphere of radius $S_0$ and integrating out the angular variables. The result reads:
\begin{equation}
\label{pdf-S0}
\begin{split}
P(S_0)&=y_1^{-1} \,S_0\, e^{\tilde \mu\,\tilde \beta S_0-\tilde \beta S_0^2/2}\,\intop_0^1\frac{dy}{\sqrt{1-y}}\,\mathrm{I}_0 \left[\tilde \beta \frac{\lambda_2}{4}\,S_0^2 \,y\right]\times \\
& \,e^{\tilde \beta S_0^2 (1-\lambda_1)y/4}.
\end{split}
\end{equation}
Here $y_1$ is the normalization constant related to the overall partition function of Eq.(\ref{Z-vector}) via $\mathcal{Z}=(2\pi y_1)^N$ (as both linear coupling and the birefringence can be neglected). The full analytical form of the intensity PDF $P(S_0)$ cannot be obtained in the general case but one can see that it is i) asymmetric and ii) has a Gaussian asymptote $P(S_0) \propto S_0^{-1}\exp[\tilde \beta \tilde \mu S_0 - (\tilde \beta/4)(1+\lambda_1-\lambda_2)S_0^2]$ as $S_0 \to \infty$. Again the condition $|\lambda_1-\lambda_2|<1$ ensures the convergence. The normalized inverse temperature $\tilde \beta$ and the chemical potential $\tilde \mu$ can then be determined in a standard way from Eqs.(\ref{h-p}). By closely inspecting the normalization integral $y_1$ one can infer that it has a self-similar form (and so does the partition function):
\[
y_1(\tilde \beta,\tilde \mu)= \frac{1}{\tilde \beta}\,F\left(\tilde \mu \sqrt{\tilde \beta}\right),
\]
where the explicit form of the function $F$ is given below. Plugging this ansatz into (\ref{h-p}) and excluding the terms containing the derivative of $\ln F$ we obtain the relation between the (normalized) chemical potential and the temperature:
\begin{equation}
\label{mu-vector}
\tilde \mu =\frac{h_0}{\Gamma}-\frac{1}{\tilde \beta}.
\end{equation}
The temperature is then to be obtained from a single transcendental equation:
\begin{equation}
\label{eq-beta}
F'(f_0\, \tilde \beta^{1/2} -1/\tilde \beta^{1/2})= 2\tilde \beta^{1/2}\,F(f_0\,\tilde \beta^{1/2}-1/\tilde \beta^{1/2}),
\end{equation}
where the derivative is taken with respect to the argument, and $f_0=h_0/\Gamma$ is determined via Eq.(\ref{def-f}). As for the function $F(x)$, it is generally only available in quadratures as
\[
F(x)=\frac{1}{2}\,\intop_0^\infty dz \, e^{x\sqrt{z}-z/2}\,\intop_0^1 \,\frac{dy}{\sqrt{1-y}}\, \mathrm{I}_0 \left[\frac{\lambda_2}{4}\,zy \right]e^{(1-\lambda_1)yz/4}.
\]

The solution of the transcendental equation (\ref{eq-beta}), $\tilde \beta= \beta \Gamma$ is universal inasmuch as it does not depend on the value of nonlinearity, $\Gamma$. For given choice of $\lambda_{1,2}$ it depends only on the average initial energy (\ref{def-f}) and as long as the latter is not very close to the minimal value $h_{min}$ (which corresponds to the universal regime considered in the previous section) the solution always exists an is in the order of 1.

\begin{figure*}[tb]
\includegraphics[width=178mm]{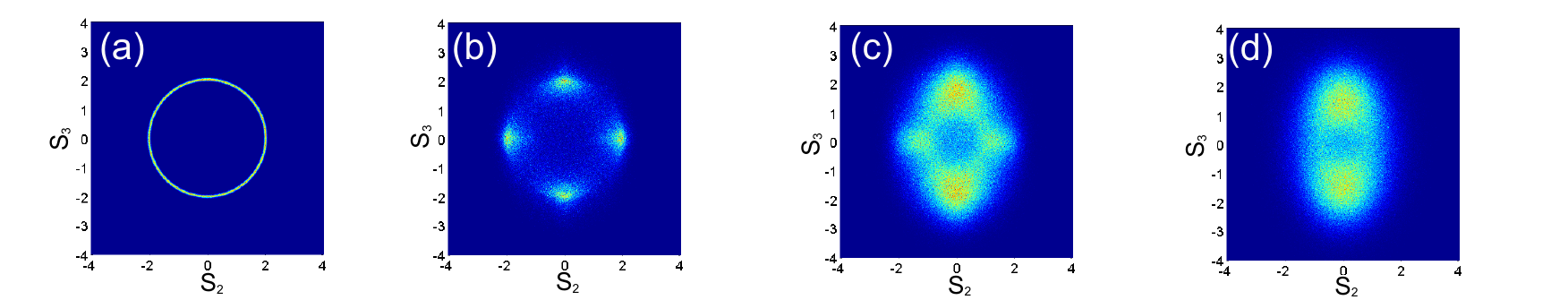}
\caption{\label{fig:evolution} (Color online) The evolution of the marginal PDF $P(S_2,S_3)$ with distance. Initial amplitudes are equal $A_0=B_0=1$ and all the phases are uniformly distributed and uncorrelated. (a) The initial uniform distribution on a circle $z=0$ (b) a 4-modal distribution is formed initially $z=1$ and is later gradually destroyed (c) at $z=9.4$ and the final 2-modal Gibbs state sets in (d) (z=50).}
\end{figure*}

In our numerical simulations shown in Fig.\ref{fig:general} we have chosen a symmetric initial condition with completely randomised phases: $A_0=B_0=1$, $\langle \cos \delta \rangle=0$. From (\ref{def-f}) this corresponds to the value $f_0=1+\lambda_1$. For the anisotropic case (AlGaAs) the numerical solution of (\ref{eq-beta}) yields $\tilde \beta = 2.32$ while for the isotropic one (fused silica) one gets  $\tilde \beta =4.41$.
We have found that unlike in the case of universal correlations observed in section \ref{sec:universal} (or the scalar case), the thermalization length now depends on the temperature (which is directly proportional to the nonlinearity). Generally, the higher is the temperature, the longer it takes for the system to reach equilibrium. For the highest level on nonlinearity achieved in our simulations ($\Gamma = 200$) the thermalization length was of the order $z_{th} \sim 500$ units of coupling length. This is of course an extreme limit and lower values of nonlinearity (i.e. temperature) produce lower values of the thermalization length (typically $~10$). In \ref{fig:general}(a)-(d) we can see marginal distribution of the 2 Stokes parameters $P(S_1,S_2)$ for $\Gamma=50$. Theoretical profiles (b),(d) were obtained by integrating out the $S_1$ variable in the (normalized) distribution (\ref{pdf-Stokes}) while the parameters $\tilde \beta$ and $\tilde \mu$ were determined via
(\ref{mu-vector}) and (\ref{eq-beta}). One can observe a good agreement between theory and numerics \cite{footnote1}. The distribution is bimodal but the peaks (corresponding to circular clockwise and anticlockwise polarisation) are wide enough (the width is in the order of $\tilde \beta^{-1/2}$). \ref{fig:general}(e),(f) compare theoretical prediction given by (\ref{pdf-S0}) with the numerics. One can see that for different values of the nonlinearity $\Gamma$ the product $\tilde \beta=\beta\,\Gamma$ (and hence the intensity PDF $P(S_0)$) indeed remains the same and instead of universal correlation functions observed in the scalar case or the case of circular initial polarization (Fig.\ref{fig:universal}(c)) one now has \textit{universal intensity distribution} as given by (\ref{pdf-S0}). The opposite case occurs with regards to the angle distribution $P(\theta)$, \ref{fig:general}(g), (h). Whereas in the the scalar case and in the case of circular polarization (Fig.\ref{fig:universal}(d)) this PDF remains invariant, now, according to the theoretical prediction (\ref{saddle-pdfs}) the distribution becomes closer and closer to uniform as $\Gamma$ increases since $h_{min}$ scales as $\Gamma$ while $\tilde \beta$ and $\tilde \mu$ remain constant. In some respect the results of (\ref{fig:general})(e)-(h) mirror those results for the circular polarization \ref{fig:universal}(c),(d) and the scalar case (Fig.2 of Ref.\cite{slbsm09}) but with the angular and intensity distribution trading places.

The universal nature of the intensity PDF $P(S_0)$ can also be explained from the energy conservation (a similar argument for the scalar case was put forward in \cite{slbsm09}). In the limit of large $\Gamma$ most of the energy concentrates in the nonlinear part $\mathcal{H}_\gamma$ which is a sum of on-site component given by (\ref{Hamiltonian-gamma}). We can introduce the standard deviation for intensities as $\sigma^2(z)=\langle (A+B)^2 \rangle -4$ (where angular brackets denote both averaging over the initial conditions and over waveguides) and rewrite the nonlinear energy in terms of $\sigma^2$. Then the following average integral of motion exists:
\begin{equation}
\label{int-mot}
\Gamma \sigma^2(z) + \Gamma \langle (\lambda_1-1) AB+\lambda_2 AB \cos(2\delta) \rangle + O[1]=const,
\end{equation}
where the term $O[1]$ collectively denotes the contribution from the linear coupling and birefringence. The value of the constant is determined by the initial conditions and since initially all waveguides have the amplitudes, $A_0$, $B_0$, $A_0+B_0=2$, and random uncorrelated phases the initial variance vanishes $\sigma^2(0)=0$ and one has the following relation:
\[
\sigma^2(z)=-\langle (\lambda_1-1)(AB-A_0B_0)+\lambda_2 AB\cos(2\delta) \rangle +O\left[1/\Gamma\right].
\]
This means that beyond the thermalization distance $z_{th}$ the first term in the r.h.s. becomes a stationary $\Gamma$-independent constant of the order of unity and so does the variance of the intensity distribution as clearly seen in \ref{fig:general}(e). The situation changes however if one starts from the locked circular polarization state for each waveguide as in section \ref{sec:universal}. Then the first term in the r.h.s. of the above changes to $-\langle (\lambda_1-1)(AB-1)+\lambda_2 AB(1+\cos(2\delta)) \rangle$. But the distribution in this case centres narrowly on the extremal values $A_*=B_*=1$, $\delta_* =\pm \pi/2$ so that the average vanishes and in this regime the variance scales inversely proportional to $\Gamma$ as is indeed seen in \ref{fig:universal}(c).

Finally it is instructive to see how the distribution of the polarization state $P(S_2,S_3)$ evolves with distance $z$, passing from the initial isotropic state through the integrable regime described in Section \ref{sec:anticont} and eventually relaxing towards the equilibrium Gibbs distribution. The successive snapshots of such evolution are given in Fig.\ref{fig:evolution} for the anisotropic case and $\Gamma=50$. One can see all successive stages from the uniform distribution on a circle of radius 2 onto the 4-modal distribution that occurs in a transient anticontinuum regime when the energy exchange between the waveguides is still negligible (cf. Fig.\ref{fig:anticont}(a)) and finally towards the thermal equilibrium state which is achieved when the weak linear inter-waveguide coupling provides uniform mixing between all the degrees of freedom.

\section{Conclusions}

In this paper we have considered the structure of a thermalized field dynamics in discrete birefringent waveguide systems. We have defined for the first time an exact boundary in the space of the integrals of motion separating the thermal phase with positive temperature from that corresponding to localized excitations (discrete breathers). We have shown that in the limit of high nonlinearity depending on the choice of the initial conditions the marginal PDF for the stokes parameters $P(S_2,S_3)$ relaxes either to a universal, broad, bimodal distribution (with the maxima corresponding to clock- or anti-clockwise circular polarization) or (if initially all waveguides are locked in the same circular polarization) a narrow-peaked one, corresponding to the thermal fluctuation around the initial state. In both cases either the effective ``temperature constant'' $\beta$ is universal (i.e. does not depend on the nonlinearity parameter $\Gamma$) or generally the product $\beta \Gamma$ remains fixed as one increases the value of the nonlinearity.

Also in the limit of strong nonlinearity before reaching the final thermal state the system passes through the state where its dynamic is integrable, corresponding to the anticontinuum limit of the nonlinear polarization dynamics. In this regime the probability distribution of finding a system in  a certain polarization state has a complex multimodal structure. As the thermalization sets in the different modes of the distribution gradually disappear leaving only the two maxima corresponding to the circularly polarized states achieving the global minima of the nonlinear coupling energy subject to fixed total power.

As a possible continuation of this research one may suggest a more detailed study of the material dependence of the final thermal state (here only two specific examples were considered) as well as a full thermodynamical analysis of the breather region of the phase space as done by Rumpf \cite{r08,r09} for the scalar case.

\begin{acknowledgments}
The author would like to thank Yaron Silberberg and Omer Sidis for illuminating discussions. This work was supported by the Marie Curie Fellowship project~``INDIGO''.
\end{acknowledgments}

\appendix

\section{The dynamics of Stokes parameters in the anticontinuum limit}
\label{sec:appendix}

In this section we recall the exact results for the dynamics of the system of Stokes parameters (\ref{Stokes-dyn}) (see \cite{contvectdnls,couplers}). By differentiating the first Eq. in (\ref{Stokes-dyn}) and  using the other two we arrive at the following system:
\begin{equation}
\label{k-0}
\begin{split}
S_1''+\omega^2\,S_1=0 & \quad \mbox{Isotropic case}\\
S_1''+\alpha \,S_1 +\beta\,S_1^3 =0 & \quad \mbox{Anisotropic case}
\end{split}
\end{equation}
with
\begin{eqnarray*}
 \alpha&=&-\frac{\Gamma^2}{2}\,\left[(3\lambda_1-2)(\lambda_1-2)S_1^2(0)+\lambda_1(2-3\lambda_1)S_2^2(0) \right. \\
 & + & \left. \lambda_1(\lambda_1-2)S_3^2(0)\right] \\
\beta&=&\frac{1}{2}\,\Gamma^2\,(2-\lambda_1)(2-3\lambda_1), \quad \omega=2\gamma(1-\lambda_1)\,S_3(0)
\end{eqnarray*}
subject to initial conditions $S_1(0)$, and $S'_1(0)=2\Gamma \lambda_2 S_2(0) S_3(0)$. The coefficient $\beta$ should not be confused with the reciprocal temperature as defined in other sections. Thus in the isotropic case we get an equation for a harmonic oscillator while in the anisotropic case the equation is that of the nonlinear Duffing oscillator.

The solution in the isotropic case is simple:
\begin{equation}
S_1(z)=S_1(0)\,\cos(\omega \,z) +S_2(0)\,\sin(\omega z).
\label{s1-isotropic}
\end{equation}
In the anisotropic case the form of the solution depends on the sign of the coefficient $\beta$ as well as the Duffing oscillator energy:
\begin{eqnarray}
\label{Duffing-Energy}
E&=&\frac{{S'}_1^2}{2}+\frac{\alpha}{2}S_1^2+\frac{\beta}{4}\,S_1^4  = \frac{1}{8} \Gamma ^2 \left(-S_1^2 (-2+\lambda_1 )+2 \lambda_1  S_2^2 \right) \nonumber \\
 &\times & \left(2 \lambda_1 S_3^2 +S_1^2 (-2+3 \lambda_1 )\right)
\end{eqnarray}
Here we will only consider the case when $2/3< \lambda_1 \leq 2$ so that we have simultaneously $\beta<0$, $\alpha>0$ and $E>0$. For AlGaAs we have $\lambda_1 =1$ which corresponds exactly to this regime. Moreover in this case the potential energy has 2 symmetric maxima with the values $E_*=-\alpha^2/4\beta$ and one can prove that for any choice of the initial conditions $\{S_1(0),S_2(0),S_3(0)\}$ the difference
\[
E-E_*=\frac{\Gamma ^2 \lambda_1^2 \left(S_3^2(0) (-2+\lambda_1 )+S_2^2(0) (-2+3 \lambda_1 )\right)^2}{8 \left(4-8 \lambda_1 +3 \lambda_1^2\right)}<0.
\]
The effective ``particle'' oscillates in the valley between the two maxima. The solution is expressed via Jacobi elliptic function:
\begin{eqnarray}
\label{Duffing-Solution}
S_1(z)&=&S_1^*\,\mathrm{sn}\left[F(S_1(0)/S_1^*,\kappa)+\sqrt{\frac{\alpha}{1+\kappa^2}} z,\kappa \right], \\
S_1^*&=&\sqrt{\frac{\alpha}{|\beta|}}\,
\frac{\sqrt{2}\kappa}{\sqrt{1+\kappa^2}},  \quad \kappa=\sqrt{\frac{E_*}{E}}-\sqrt{\frac{E_*}{E}-1} \leq 1. \nonumber
\end{eqnarray}
Here $S_1^*$ is the maximal reachable amplitude of the oscillations, $\mathrm{sn}(x,\kappa)$ is the elliptic sine function and $F(x,\kappa)$ is the incomplete elliptic integral of the first kind:
\[
F(x,\kappa)=\intop_0^x \,\frac{dt}{\sqrt{(1-t^2)\,(1-\kappa^2t^2)}}.
\]
Now using either the two integrals of motion of the main system  - i.e. Eqs(\ref{Stokes-integrals}) or integrating the system (\ref{Stokes-dyn}) directly one can restore the remaining Stokes parameters, $S_2(z)$ and $S_3(z)$. In the anti-continuum limit the averaging over the initial conditions amounts to averaging either (\ref{s1-isotropic}) or (\ref{Duffing-Solution}) for $S_1$ and using the integrals of motion to obtain the marginal PDF $P(S_2,S_3)$.


\begin{thebibliography}{99}
\bibitem{footnote1} Becasue of the Jacobian $(2S_0)^{-1}$ there is a narrow integrable singularity at zero both in (\ref{pdf-Stokes}) and in the marginal PDF $P(S_2,S_3)$, but the width of this singularity is much smaller than the computational grid and it does not feature prominently in our simulations.
\end{thebibliography}
\end{document}